# Controlling the anisotropy of a van der Waals antiferromagnet with light

D. Afanasiev[1], J.R. Hortensius[1], M. Matthiesen[1], S. Mañas-Valero[4], M. Šiškins[1], M. Lee[1], E. Lesne[1], H.S.J. van der Zant[1], P.G. Steeneken[1], B.A. Ivanov[2,3], E. Coronado[4] and A.D. Caviglia[1]

[1]Kavli Institute of Nanoscience, Delft University of Technology, P.O. Box 5046, 2600 GA Delft, The Netherlands.
[2] Institute of Magnetism, National Academy of Sciences and Ministry of Education and Science, 03142 Kyiv, Ukraine.
[3]National University of Science and Technology «MISiS», Moscow, 119049, Russian Federation.
[4]Instituto de Ciencia Molecular (ICMol), Universitat de Valencia Catedrático José Beltrán 2, 46980 Paterna, Spain

**Abstract:** Magnetic van der Waals materials provide an ideal playground for exploring the fundamentals of low-dimensional magnetism and open new opportunities for ultrathin spin-processing devices. The Mermin-Wagner theorem dictates that as in reduced dimensions isotropic spin interactions cannot retain long-range correlations; the order is stabilized by magnetic anisotropy. Here, using ultrashort pulses of light, we demonstrate all-optical control of magnetic anisotropy in the two-dimensional van der Waals antiferromagnet $NiPS_3$. Tuning the photon energy in resonance with an orbital transition between crystal-field split levels of the nickel ions, we demonstrate the selective activation of a sub-THz two-dimensional magnon mode. The pump polarization control of the magnon amplitude confirms that the activation is governed by the instantaneous magnetic anisotropy axis emergent in response to photoexcitation of orbital states with a lowered symmetry. Our results establish pumping of orbital resonances as a universal route for manipulating magnetic order in low-dimensional (anti)ferromagnets.

The recent discoveries of van der Waals (vdW) two-dimensional (2D) layered magnets have led to a surge of interest due to their potential applications in constructing atomically-thin spin-processing devices and non-volatile magnetic memories[1,2]. Unique phenomena and effects are foreseen in 2D magnetic systems due to their reduced dimensionality[3-5]. In contrast to three-dimensional magnets, long-range magnetic order cannot exist in two dimensions at any finite temperature without the presence of magnetic anisotropy[6,7]. In 2D magnets, the anisotropy not only sets a preferred direction for spins but also protects the magnetic order against dimensionality-enhanced thermal spin fluctuations. This intimate relationship between magnetic order and anisotropy in 2D motivates the ongoing search for efficient pathways to manipulate the magnetic anisotropy in such systems. As the magnetic anisotropy in most materials is determined by the coupling of electronic orbitals and spins, stabilizing and controlling 2D magnetism is actively pursued through the manipulation of orbital degrees of freedom, using, for example, mechanical strain[8-10] and electrostatic gating[11,12]. However, a large anisotropy is normally associated with an unquenched orbital moment, which is limited to specific oxidation states and to low-symmetry crystal environments, most notably for rare-earth ions[13]. In most 2D magnets, magnetism arises from transition metal ions, which typically have a quenched orbital moment in their ground state. In these systems magnetic anisotropy arises due to the spin-orbit driven mixing of the ground state with higher-energy orbital states with unquenched momentum, a rather small effect. Optical pumping of the electronic transition towards the higher-level orbital states (orbital resonances) provides the most direct access to the admixing and subsequent control of the magnetic anisotropy as manifested by the excitation of spin precession in 3D magnets[14-16] even to the extent of the sub-cycle coherent switching of the spin orientation[17-19]. Two-dimensional magnets, with their subtle interplay between anisotropy and magnetic order, offer an intriguing playground for exploring the effect of resonant pumping of orbital transitions on the magnetic anisotropy.

Here we study optical control of magnetism in nickel phosphorus trisulfide $NiPS_3$, a novel van der Waals layered magnet with XY-type antiferromagnetism[20-22]. The energetic competition between charge-transfer and Coulomb repulsion makes this system an intriguing example of a strongly correlated 2D magnet, with pronounced spin-charge correlations[23], spin-orbit entangled excitons[24], and strong spin-lattice coupling[25]. We optically pump $NiPS_3$ using ultrashort pulses of light and probe the ensuing spin dynamics on the picosecond timescale. Continuously varying the pump photon energy across orbital resonances, we focus on a transition to the orbital state responsible for the anisotropic magnetic properties in equilibrium and demonstrate the selective activation of a hitherto unreported sub-THz magnon mode with a pronounced 2D character. By studying the mode's excitation as a function of the pump polarization and photon energy, we show that the activation indeed proceeds as a result of the instantaneous magnetic anisotropy due to the low-symmetry nature of the photoexcited orbital states. We also find that optical pumping in the region of optical transparency impulsively activates another high-frequency coherent mode, a previously unreported candidate for the complementary magnon mode in $NiPS_3$.

$NiPS_3$ crystallizes in the monoclinic space group $C/2m$[26], see Fig. 1a. In the *ab* plane, it features a network of edge-sharing $NiS_6$ octahedra arranged on a honeycomb lattice, each having a small trigonal distortion perpendicular to this plane, see Fig. 1b. Below the Néel temperature $T_N$=155 K, the magnetic moments of $Ni^{2+}$ ions arrange into a complex compensated antiferromagnetic pattern. The pattern is formed by zig-zag ferromagnetic spin chains along the *a*-axis, which are coupled antiferromagnetically within the single layer[27], see Fig. 1a. A large spacing $c$=6.63 Å between adjacent layers leads to a negligible orbital overlap between the magnetic ions of different layers, thereby suppressing interlayer exchange such that the antiferromagnetic order acquires a 2D character already in the bulk form.

The orientation of magnetic moments in $NiPS_3$ is governed by a biaxial magneto-crystalline anisotropy consisting of two distinct contributions: a dominant easy-plane anisotropy which locks the orientation of the spins to a magnetic plane *(xy)*, slightly inclined from the crystallographic *ab*-plane; and a secondary weaker anisotropy which orients the spins in the magnetic *(xy)*-plane along the *x*-axis. Microscopically, the easy-plane anisotropy develops as a result of a zero-field splitting ($D \approx -1.1$ meV[28]) of the $^3A_{2g}$ ground state of the $Ni^{2+}$ ion ($S = 1$) in the crystal field of the trigonally distorted $NiS_6$ octahedra, see Fig. 1b. Note that $^3A_{2g}$ is an orbital singlet and alone cannot develop the splitting. The splitting and anisotropy arise indirectly as a consequence of spin-orbit driven intermixing of the ground state with the first excited orbital triplet state $^3T_{2g}$, which is split by the trigonal lattice distortion[28,29] into a set of low-symmetry doublets $^3E_g$ and singlets $^3A_{1g}$ separated by an energy gap of around 110 meV, as schematically shown in Fig. 1b. Although there are no reports on the origin of the in-plane magnetic anisotropy along the *x*-axis in $NiPS_3$, it likely stems from a rhombic distortion of the $NiPS_6$ octahedra, which further splits the $^3A_{2g}$ levels. Hence, an anisotropic Hamiltonian considering not only the axial distortion of the octahedron, but also an in-plane distortion may be needed to take this observation into account (SI5).

The orbital resonances in $NiPS_3$ correspond to a pair of *d-d* transitions $^3A_{2g} \rightarrow {}^3T_{2g}$ and $^3A_{2g} \rightarrow {}^3T_{1g}$ emerging within the $^3F$ ground state multiplet of the $Ni^{2+}$ ion split by the octahedral crystal field ($O_h$) (see SI1). In $NiPS_3$ these transitions result in a pair of two broad absorption bands centred at 1.07 eV ($^3A_{2g} \rightarrow {}^3T_{2g}$) and 1.73 eV ($^3A_{2g} \rightarrow {}^3T_{1g}$). Note that in contrast to other transition metal ions, the *d-d* resonances in $Ni^{2+}$ are spin-parity allowed ($\Delta S$=0), i.e. they do not involve a spin-flip, and thus cannot directly affect the exchange interaction between adjacent spins. To selectively address these resonances, we employed ultrashort (~100 fs) pump pulses with photon energy tunable in a broad spectral range of 0.1 eV-1.9 eV. The pump-induced dynamics were measured by tracking the intensity $I$ and the rotation of the polarization plane $\theta$ of time-delayed co-propagating near-infrared probe pulses at a photon energy of 1.55 eV, as schematically shown in Fig. 1c. Whereas $I$ is a measure of the non-magnetic diagonal components of the dielectric tensor, $\theta$ is proportional to the off-diagonal components sensitive to the magnetic order via magneto-optical effects, e.g. the Faraday effect and/or magnetic linear birefringence.

The sample was cooled down to 10 K well below $T_N$ and pumped using linearly polarized pulses at variable photon energies. The time-resolved magneto-optical rotation $\theta$ reveals a striking sensitivity of the pump-induced dynamics to the photon energy of the excitation (see Fig. 1d,e). When excited at the $^3A_{2g} \rightarrow {}^3T_{2g}$ resonance, $\theta$ displays a damped oscillation as a function of the pump-probe time delay $\Delta t$, with frequency $f_1$=0.30 THz (see SI2 for the Fourier spectra). No coherent oscillations were observed when exciting at the higher photon energy corresponding to the $^3A_{2g} \rightarrow {}^3T_{1g}$ resonance. Detuning the photon energy below the absorption lines of the resonances ($hv$=0.8 eV) shows no signal associated with the frequency $f_1$, but reveals instead another higher-frequency mode at $f_2$=0.92 THz. We note that the oscillations seen in the polarization rotation dynamics were not observed in the probe intensity dynamics $I$ (see SI3), thus hinting at their magnetic origin. In addition, we found no match for the frequencies of these oscillations in the phonon spectrum of NiPS$_3$, well studied in recent years[25,30-32].

To understand the significance of the orbital resonances, we tracked the ultrafast dynamics while varying the pump photon energies across the subgap states down to the phonon Reststrahlen band edge at 0.1 eV. The amplitudes of both oscillations at $f_{1,2}$ were retrieved and their relative values (normalized on the pump fluence) plotted as a function of the pump photon energy. Figure 2a shows that the $f_2$ mode is excited with almost equal efficiency between 0.1 and 0.9 eV, in the optical transparency window, where optical absorption is negligible. Nearly complete suppression of the oscillation amplitude occurs in proximity to all absorption lines, including the low-energy lattice absorptions below 0.1 eV[33]. The fact that the $f_2$ mode is excited in the transparency window implies that it is initiated by a mechanism relying on virtual electronic transitions, i.e., due to an off-resonant impulsive Raman scattering process (ISRS)[34,35]. The suppression of the excitation at all other photon energies is explained by a decrease in the scattering efficiency as other absorption channels start to compete.

In striking contrast, the excitation of the lower-frequency $f_1$ mode only occurs in a relatively narrow photon energy range, showing a pronounced resonance with the $^3A_{2g} \rightarrow {^3T_{2g}}$ transitions, see Fig. 2a. The lineshape of the resonance reveals a fine structure indicative of the trigonal splitting of the $^3T_{2g}$ manifold ($^3E_g$, $^3A_{1g}$), see SI1. Despite a nearly order of magnitude stronger optical absorption, no oscillations were seen upon resonant pumping of the $^3A_{2g} \rightarrow {^3T_{1g}}$ higher energy orbital resonance, underscoring the exceptional sensitivity of the oscillations to the photoexcitation of the $^3T_{2g}$ states. Remarkably, whereas the amplitude of the mode at $f_2$ reveals a linear dependence on the pump fluence (Fig. 2b), the amplitude of the $f_1$ mode saturates above 5 mJ/cm$^2$ (Fig. 2c), indicating a possible saturation of the $^3A_{2g} \rightarrow {^3T_{2g}}$ transition.

The temperature ($T$) dependence of the frequencies $f_{1,2}$ evidences that these modes are indeed sensitive to the magnetic ordering. Figure 3a shows that as $T$ increases, the damping of the first mode goes up while the frequency $f_1$ gradually decreases and ultimately converges to zero at a temperature close to $T_N$. Although the application of a relatively weak in-plane magnetic field $H$ up to 7 kOe produced no observable shift in $f_1$ (see SI4), the observation of the critical softening is a strong indication that the oscillation is of magnetic origin[36,37]. The softening can be characterized by a power law $f_1(T) \propto (T_N - T)^\beta$ (see Fig. 3c), with $T_N \approx 155 \pm 1$ K, in full agreement with literature data, and a critical exponent $\beta$=0.22±0.02. Note that this $\beta$ value also matches remarkably well with the critical exponent of the XY-model ($\beta_{XY}$=0.23) previously proposed to describe the temperature evolution of the 2D magnetic ordering in NiPS$_3$[38]. This remarkable observation is an unambiguous indication of the intrinsically 2D character of the mode.

The temperature-evolution of the higher-frequency oscillation at $f_2$ is significantly different. As $T$ increases, the central frequency $f_2$ shows a slight increase, which above 75 K is followed by a steep, nearly linear, softening. A linear extrapolation of the frequency decrease versus $T$ suggests that a complete softening of the mode occurs at $T$=170 K, in proximity to $T_N$ (see Fig. 3d). The softening indicates that the oscillation is either of magnetic origin itself or strongly sensitive to the magnetic ordering. This is further corroborated by the significant growth of the damping constant upon heating. Such highly damped behaviour is typical for soft modes in the vicinity of their associated phase transitions[39].

We now analyze the spin dynamics possible in NiPS$_3$ from a phenomenological theory perspective. Two magnon modes are expected in a compensated antiferromagnet, featuring a biaxial magnetic anisotropy[40,41]. The modes correspond to orthogonal deflections of the Néel vector defined as $\boldsymbol{L} = S(\boldsymbol{S}_1 - \boldsymbol{S}_2)$, where $S = S(T)$ is the average value of the Ni$^{2+}$ spin and $\boldsymbol{S}_{1,2}$ is a pair of antiferromagnetically coupled spins. In equilibrium, $\boldsymbol{L}$ is oriented along the $x$-axis, and deflections are expected in ($\parallel$) and out of ($\perp$) the magnetic easy-plane ($xy$), in such a way that the dynamical components $\Delta L_y$ and $\Delta L_z$ emerge (see Fig. 3e and Fig. 3f). The frequencies $f_{\parallel,\perp}$ of the magnons are defined by the geometric mean of the respective magnetic anisotropy ($D_{\parallel,\perp}$) and exchange energy $J_{ex}$ (see SI5) and in addition proportional to $S(T)$. Hence, both should experience a power law temperature scaling inherent to the magnetic order parameter $\boldsymbol{L}$ similarly to the one observed for the $f_1$ mode. As the out-of-plane anisotropy is typically more substantial for easy-plane antiferromagnets such as NiPS$_3$, $f_\parallel \ll f_\perp$ is expected. Note that although there is no net magnetization in the ground state: $\boldsymbol{M} = S(\boldsymbol{S}_1 + \boldsymbol{S}_2) = 0$, a finite magnetization component $\boldsymbol{M} \propto [\dot{\boldsymbol{L}}, \boldsymbol{L}]$ emerges due to the dynamics of the Néel vector $\boldsymbol{L}$[42]. As a consequence, the in- and out-of-plane magnetic modes can be fully described by the orthogonal pairs ($L_y$, $M_z$) and ($L_z$, $M_y$), respectively.

In Ref.[43] it was recently shown that the application of an in-plane magnetic field larger than $H_{sf}$=100 kG promotes a spin-flop transition in NiPS$_3$ during which the spins suddenly rotate in the easy-plane and in addition cant along the field orientation. It can be easily shown (SI6) that the magnitude of the spin-flop field $H_{sf}$ is a direct measure of the frequency of the in-plane dynamics $f_\parallel = \gamma H_{sf}$=280 GHz, where $\gamma$=28·10$^{-4}$ GHz/G is the gyromagnetic ratio. This estimate agrees particularly well with $f_1$ and thus provides another strong indication that the coherent oscillation excited upon resonant pumping of the $^3A_{2g} \rightarrow {^3T_{2g}}$ transition is the in-plane magnon mode characterized by $L_y$ and $M_z$ and a two-dimensional critical scaling.

Having identified $f_\parallel = f_1$, we put forward the assumption that the higher frequency oscillation at $f_2$ can be assigned to the complementary out-of-plane magnon ($f_\perp = f_2$). Indeed, our phenomenological theory (SI7) suggests that excitation of the out-of-plane magnon mode with linearly polarized light is possible in NiPS$_3$ due to the low-symmetry (monoclinic) distortion of the crystal lattice. However, these assumptions do not agree with the recently reported, although mutually conflicting, values for the zone-center magnon at the significantly higher frequencies of 1.69 and 2.4 THz from Ref.[44] and [24] respectively. To unambiguously establish the origin of the coherent mode $f_2$, time-resolved measurements in high magnetic fields $H \geq H_{sf}$, are of primary importance.

To further our understanding of the excitation mechanism of the in-plane ($f_1$) magnon and its relation to the light-induced magnetic anisotropy, we varied the orientation of the pump polarization plane, set by the azimuthal angle $\phi$ (see Fig. 4a). Although the optical absorption at the $^3A_{2g} \rightarrow {}^3T_{2g}$ orbital resonance is nearly independent of $\phi$, the amplitude and phase of the induced magnetic oscillations are strongly affected by variation of the angle. Figure 4b shows that the amplitude of the excited magnon follows a clear π-periodic sinusoid with maxima corresponding to the polarization oriented at ±45 degrees with respect to the orientation of the Néel vector *L*. This dependency can be simply understood: the linearly polarized light incident at normal to the *(ab)*-crystal plane sets up an instantaneous (within the duration of the pump pulse) magnetic anisotropy axis, directed along the orientation of the pump polarization plane. The axis breaks the magnetic symmetry in the basal plane (*xy),* providing an ultrashort in-plane magnetic torque sufficient to impulsively trigger the planar motion of the spins, see Fig. 4c. The validity of this scenario is further supported by a phenomenological theory based on symmetry considerations and general principles of light-matter interactions in a magnetic medium (SI7).

The observed polarization dependence is reminiscent of the inverse Cotton-Mouton effect[45], a particular variant of the magnetic ISRS also widely seen as a transient photo-induced magnetic anisotropy[46,47]. In the magnetic ISRS scenario, the pump photon is scattered in an event wherein an electron momentarily gains orbital angular momentum from the higher-level orbital states typically having energy higher than the incident photon. Our experimental results can be indicative of the resonant enhancement of the scattering process upon approaching the transition to the higher-level $^3T_{2g}$ orbital state, underlining the strong impact of the low-symmetry trigonal $^3T_{2g}$ states on the magnetic anisotropy of NiPS$_3$. Remarkably, our experiment shows that almost no magnon excitation is observed for the photon energies below the $^3A_{2g} \rightarrow {}^3T_{2g}$ resonance, contrary to what is expected for the ISRS process. The resonant enhancement picture must thus assume that the magnon amplitude falls below our detection limit away from the resonance. Alternatively, the excitation of spin dynamics might also result directly from changes in the magnetic anisotropy due to absorption of the photons and optical population of the low-symmetry orbital $^3T_{2g}$ states, which agrees with the observed saturation of the magnon amplitude at large pump fluences (see Fig. 2c), typical for ultrafast photo-induced anisotropy phenomena.

In conclusion, our work establishes selective pumping of orbital resonances as an efficient pathway to control magnetic anisotropy and activate high-frequency coherent spin dynamics in van der Waals layered antiferromagnets. While ultrafast control of the magnetic anisotropy and sub-THz spin dynamics is demonstrated here in bulk lamellar NiPS$_3$, due to advances in exfoliating techniques and strong magneto-optical responses observed in our experiments, we anticipate the applicability of the suggested approach to atomically thin AFMs[25,48]. Such systems can serve as an excellent testbed for the theoretical XY-model, with further possibilities such as revealing dynamics of nontrivial topological vortex states associated with the Berezinskii-Kosterlitz-Thouless transition[3,25].

## Methods:

Crystals of NiPS$_3$ were grown by chemical vapor transport. First, polycrystalline NiPS$_3$ was synthesized by mixing powders of S (99.998%, from Sigma-Aldrich), P (> 99.99%, from Sigma-Aldrich) and Ni (99.99%, from Sigma-Aldrich) in a stoichiometric ratio, pressed into a pellet and sealed in a quartz ampoule (P ~5·10$^{-5}$ mbar, length = 25 cm, internal diameter = 1.5 cm). The ampoule was kept at 400 ºC for twenty days and cooled down slowly (0.07 °C/min). Next, the previous material was mixed with iodine (99.999% anhydrous beads from Sigma-Aldrich; [I$_2$] ~ 5 mg/cm$^3$)), sealed in an evacuated quartz ampoule (P ~ 5·10$^{-5}$ mbar, length = 50 cm, internal diameter = 1.5 cm) and placed in a three-zone furnace in a gradient of temperatures of 700 °C/650 °C/675 °C for 28 days. Phase and compositional purity were verified by powder X-ray diffraction and ICP-OES (Inductively Coupled Plasma - Optical Emission Spectrometry). The materials were handled inside an argon glove-box to avoid any possible oxidation. Exact details about temperature gradients and characterization of crystals from the same batch as the ones employed in this work can be found in Ref.[8].

The pump pulses at the photon energies of 0.8-1.9 eV (~100 fs) were obtained using an optical parametric amplifier (OPA), and to access energies below 0.4 eV (~200 fs) we used difference frequency generation (DFG) by mixing the outputs of two OPAs in a GaSe crystal[49]. The pump pulses at a 500 Hz repetition rate were focused on the sample surface to a spot with a diameter of 200 µm. The time-delayed co-propagating near-infrared probe pulses at a photon energy of 1.55 eV were focused to a spot of 130 µm such that the spatial overlap between pump and probe pulses was satisfied.

## Acknowledgments:

The authors thank A.V. Kimel for critically reading the manuscript and R. Huber for continuous support.

## Funding:

This work was supported by the EU through the European Research Council, Grants No. 677458 (AlterMateria) and 788222 (MOL-2D), and the COST action MOLSPIN CA15128, the Netherlands Organization for Scientific Research (NWO/OCW) as part of the Frontiers of Nanoscience program (NanoFront), and VENI-VIDI-VICI program, the Spanish MICINN (Project MAT-2017-89993-R and Unit of Excellence "Maria de Maeztu" CEX2019-000919-M). M.Š., M.L., H.S.J.v.d.Z., and P.G.S. acknowledge funding from the EU Horizon 2020 research and innovation program under grant agreement number 785219 and 881603. E.L. acknowledges funding from the EU Horizon 2020 research and innovation programme under the Marie Skłodowska-Curie grant agreement No 707404.

## Author contributions:

D.A. conceived the project together with E.C. and A.D.C. D.A., J.R.H., and M.M. carried out the experiments and analysed the data. S.M.-V. synthesized the crystal sample. B.A.I. developed the theoretical formalism to describe the magnetization dynamics and contributed to the interpretation of the data. All authors discussed the results. The manuscript was written by D.A., J.R.H., M.M., B.A.I and A.D.C. with feedback from all co-authors.

**Competing interests:**

The authors declare no competing interests.

**Data availability:**

All the data are available on reasonable request

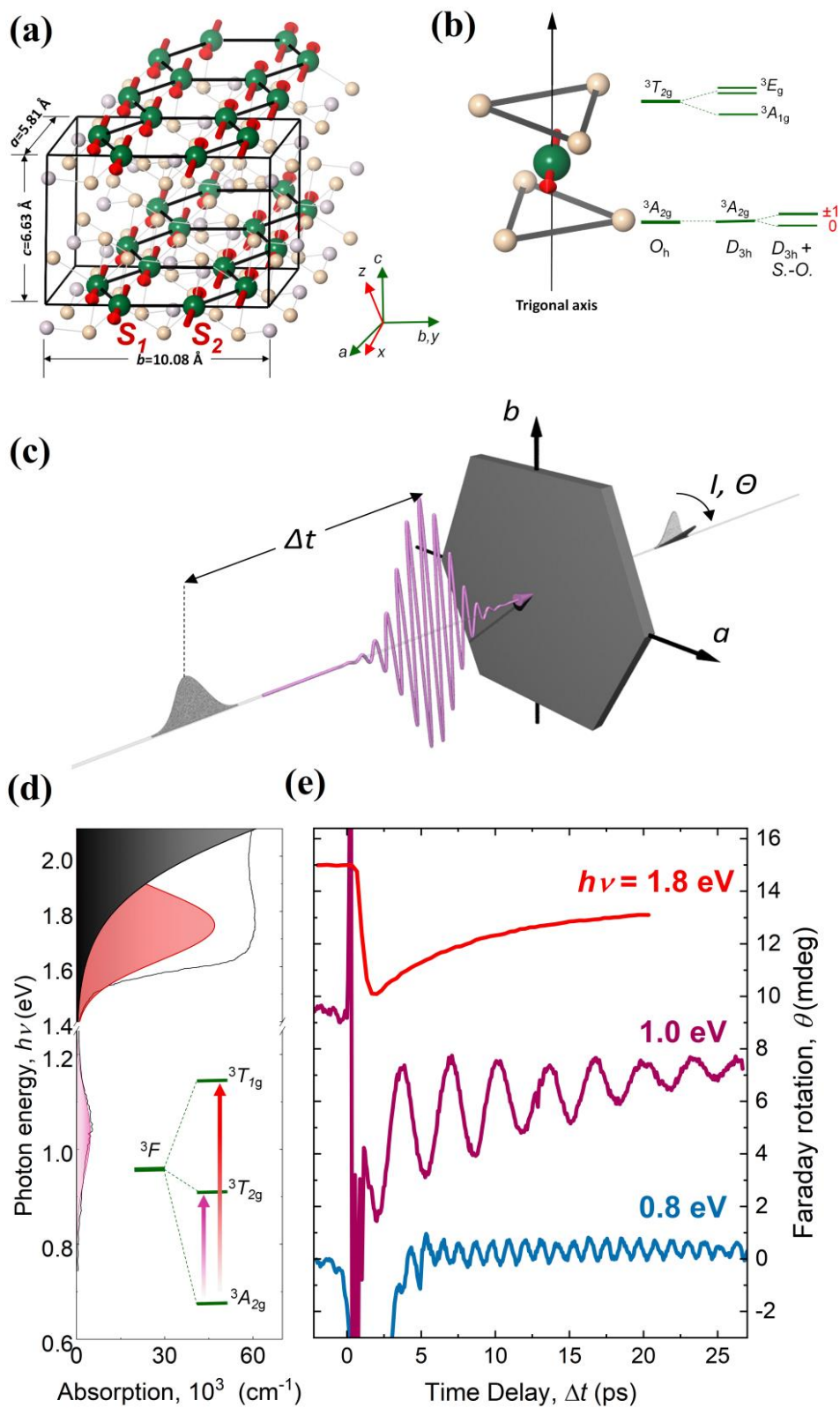

**Figure 1. Ultrafast light-induced dynamics in van der Waals antiferromagnet NiPS₃.** (a) Crystallographic and magnetic structure of NiPS₃. Green/fade orange/light pink spheres represent nickel/sulphur/phosphorous atoms. The green and red triple vectors are crystallographic and magnetic frames, respectively. **(b)** Left panel**:** $Ni^{2+}$ ion in the trigonally distorted octahedral sulphide environment**.** Right panel: Crystal field splitting of the ground state and first excited triplet state for $Ni^{2+}$ ion ($O_h$: octahedral field) in a trigonally distorted octahedral environment ($D_{3h}$); S.-O.: spin-orbit coupling **(c)** Schematic of the time-resolved pump-probe experiment. The pump (pink) and near-infrared probe pulses (grey) are collinearly focused onto the $NiPS_3$ sample with variable time delay $\Delta t$. The pump induced dynamics is measured by tracking the pump-induced polarization rotation $\theta$ and intensity $I$ of the probe pulses. **(d)** Optical absorption spectrum of $NiPS_3$ displaying $^3A_{2g} \rightarrow ^3T_{2g}$ and $^3A_{2g} \rightarrow ^3T_{1g}$ absorption bands due to the *d-d* orbital resonances of $Ni^{2+}$ ions (in pink and red respectively), and the onset of the above bandgap absorption due to Ni-S charge transfer transitions (black band). **(e)** Experimentally detected polarization rotation signal $\theta$ as a function of the delay time $\Delta t$, after excitation with pump pulses at various photon energies.

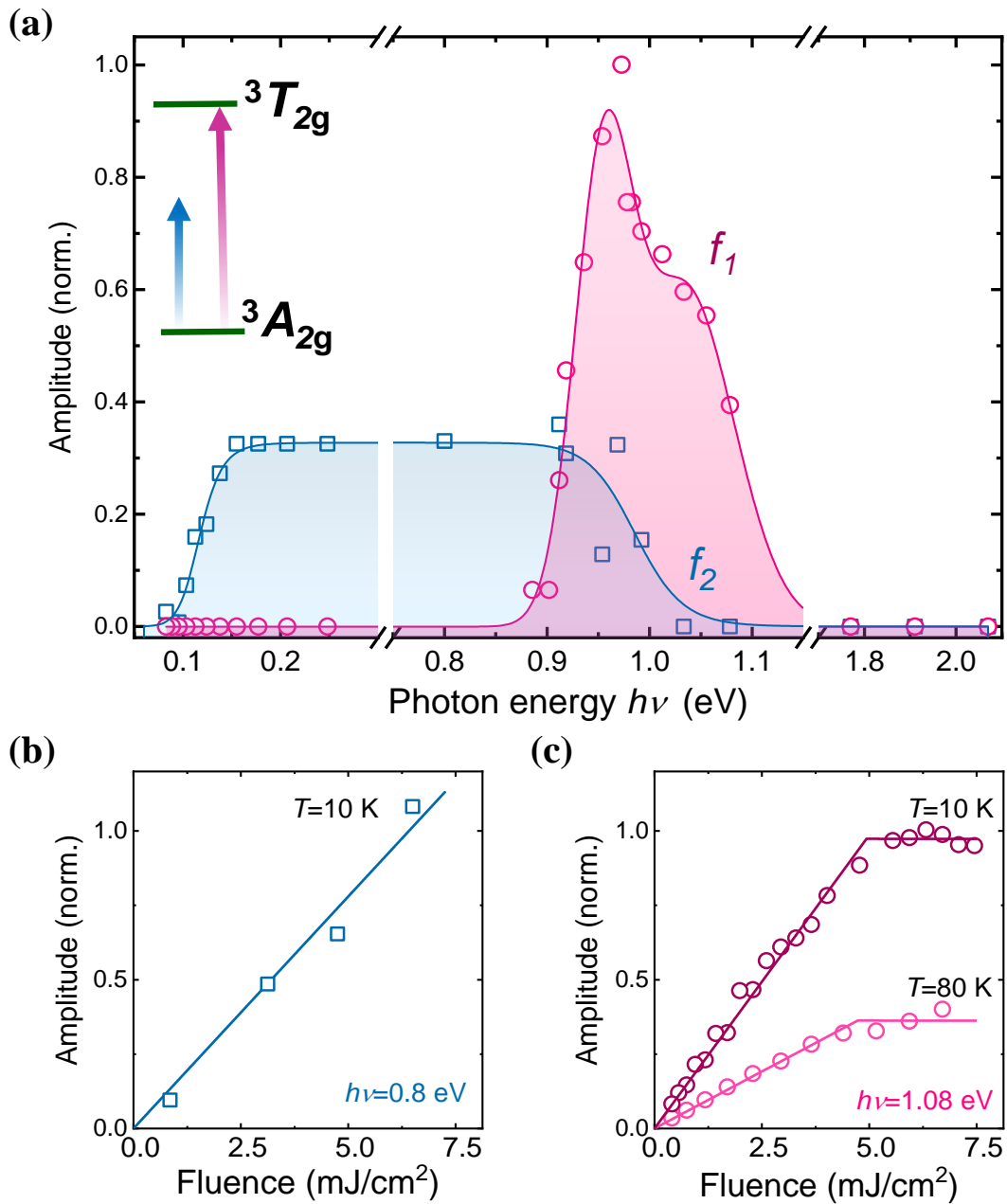

**Figure 2. Selective excitation of the light-induced coherent dynamics.** (**a**) Amplitudes of the coherent oscillations corresponding to the modes at $f_{1,2}$ normalized to the maximal value of the $f_1$ mode as a function of the pump photon energy. Solid lines are guides to the eye. Inset: Schematic illustration of the optical transition at which the $f_1$ mode is observed. (**b,c**) Amplitude of the oscillation as a function of the pump fluence for the (b) $f_2$ and (c) $f_1$ mode. The solid lines are linear fits, including saturation.

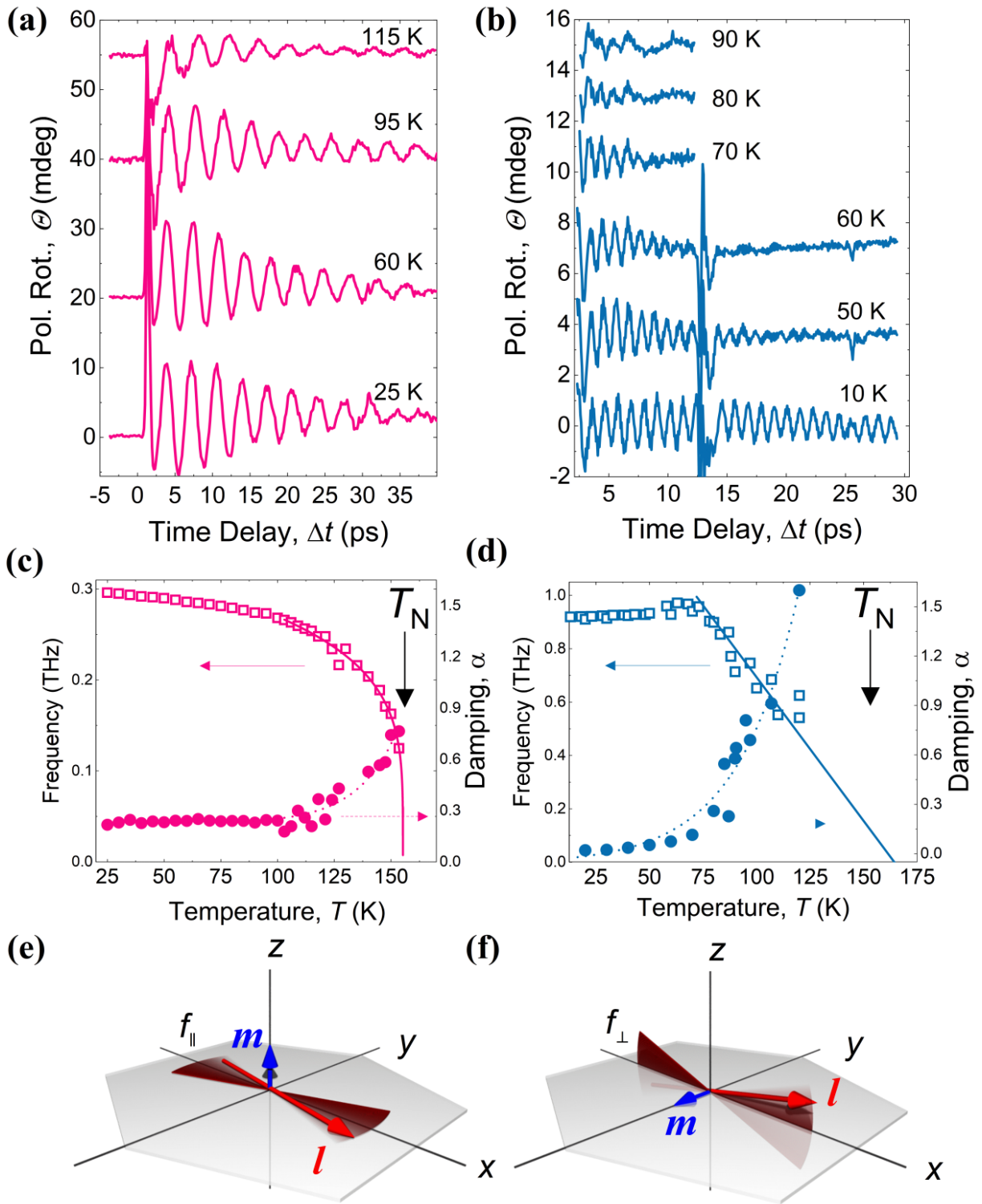

**Figure 3. Critical behaviour of the light-induced coherent dynamics.** (**a**) Temperature dependence of the time-resolved polarization rotation after pump-induced excitation with photon energies of 1.08 eV (a) and 0.2 eV (b). Irregularities at ~13 ps in (b) are related to an artefact of the measurement setup. (**c,d**) Frequency (left axis) and damping factor α (right axis) of the oscillations as extracted from the damped sine fits of the time-domain data in (a), (b). (**e,f**) Schematics of (e) the in-plane and (f) out-of-plane magnon modes.

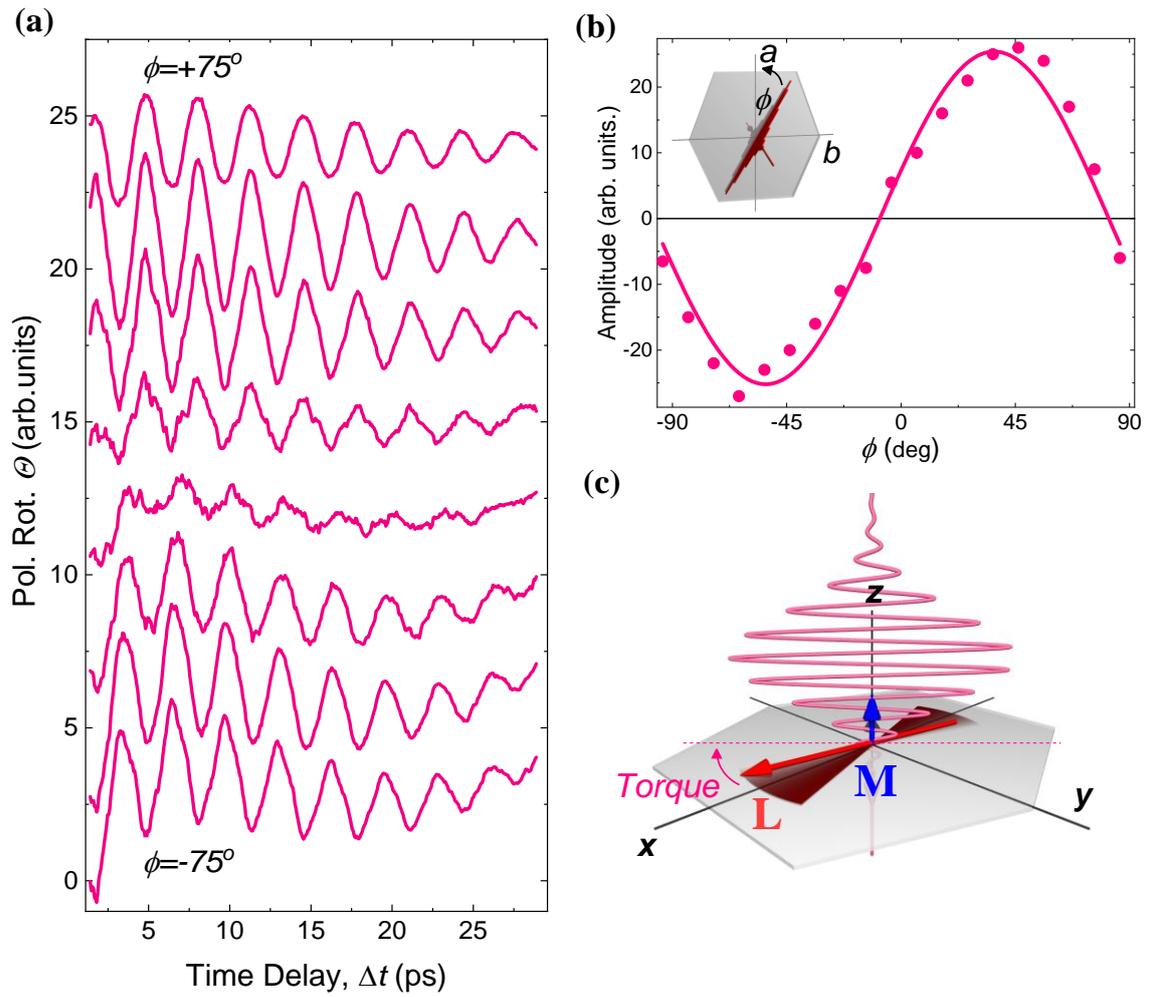

**Figure 4. Selection rules for the excitation of the in-plane magnetic mode.** (a) Pump-induced rotation of the probe polarization plane as a function of pump-probe delay time $\Delta t$ for different orientations of the linear polarization of the pump. The pump photon energy is 1.0 eV (b) Amplitude of the oscillations as a function of the azimuthal angle $\phi$ between the pump polarization plane and the *a*-crystal axis. (c) Schematics showing that the electric field of light acts as an instantaneous photo-magnetic anisotropy (dashed line) with a direction along the light polarization plane, resulting in torque and subsequent oscillations of the Néel vector **L** and net magnetization **M**

# Controlling the anisotropy of a van der Waals antiferromagnet with light

# Supplementary Information

D. Afanasiev[1], J.R. Hortensius[1], M. Matthiesen[1], S. Mañas-Valero[4], M. Šiškins[1], M. Lee[1], E. Lesne[1], H.S.J. van der Zant[1], P.G. Steeneken[1], B.A. Ivanov[2,3], E. Coronado[4] and A.D. Caviglia[1]

[1]*Kavli Institute of Nanoscience, Delft University of Technology, P.O. Box 5046, 2600 GA Delft, The Netherlands.*
[2]*Institute of Magnetism, National Academy of Sciences and Ministry of Education and Science, 03142 Kiev, Ukraine.*
[3]*National University of Science and Technology «MISiS», Moscow, 119049, Russian Federation.*
[4]*Instituto de Ciencia Molecular (ICMol), Universitat de Valencia, Catedrático José Beltrán 2, 46980 Paterna, Spain*

1. **The splitting of the ground state levels of Ni$^{2+}$ by crystal field**

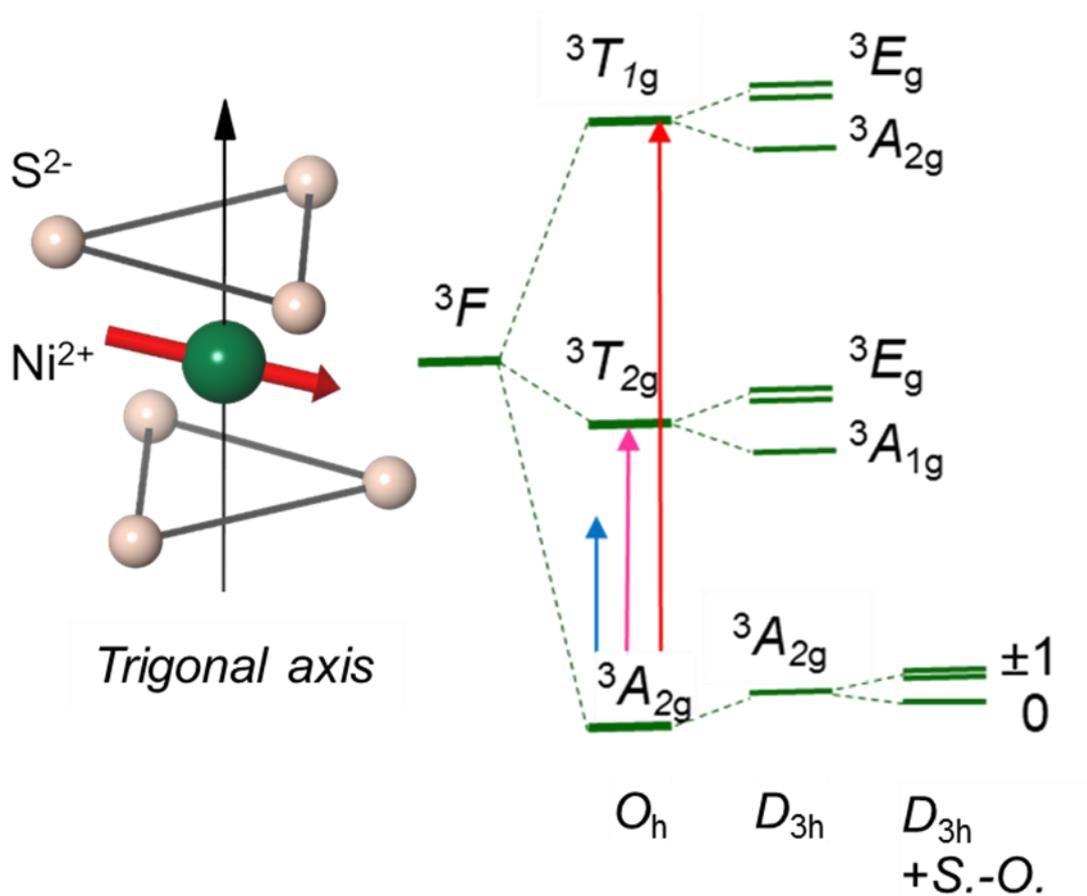

**Figure S1.** The splitting of the ground levels of Ni$^{2+}$ by octahedral ($O_h$) and trigonal fields ($D_{3h}$). The additional splitting of the ground state singlet $^3A_{2g}$ is due to the combined action of the trigonal crystal field and the spin-orbit coupling. (0, ±1) correspond to the projection of the spin moment along the trigonal distortion axis. The blue, pink and red lines symbolize the photon energies of the optical excitation used in the experiment shown in Fig. 1.

## 2. Fast Fourier Transform (FFT) analysis of the pump-induced traces

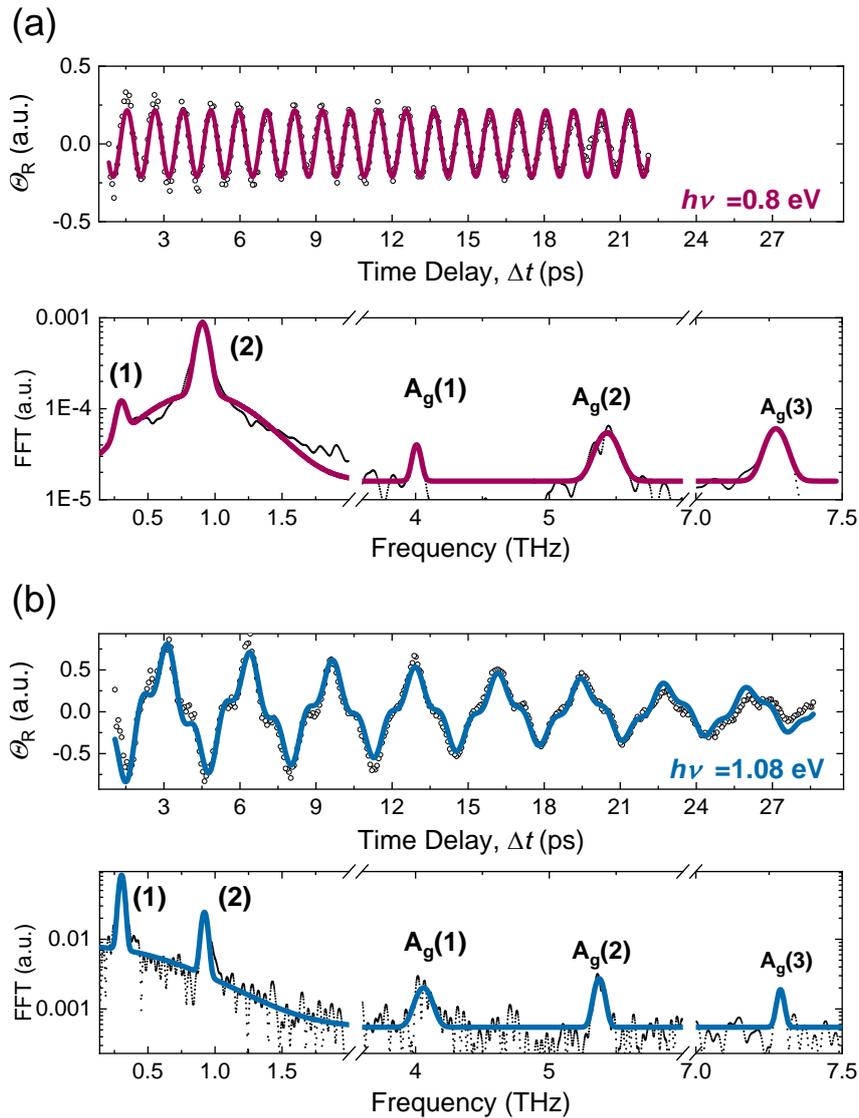

**Figure S2. (a,b)** Transient polarization rotation $\theta_R$ as a function of the delay time $\Delta t$ together with the corresponding FFT spectrum following excitation with the pump pulses at the photon energy of 0.8 eV (a) and 1.08 eV (b). Aside from the magnetic modes denoted (1) and (2), the FFT spectrum reveals a set of high-frequency $A_g$ Raman-active phonon modes previously reported for NiPS$_3$, see Refs. *Nat. Commun.*, **10**(1), 1-9 (2019); *Nanotechnology,* **30**(45), 452001 (2019); *Sci. Rep.*, **6,** 20904 (2016); *J. Phys. Chem. C* **48,** 27207 (2017).

## 3. Transient transmission at various photon energies of the pump pulse

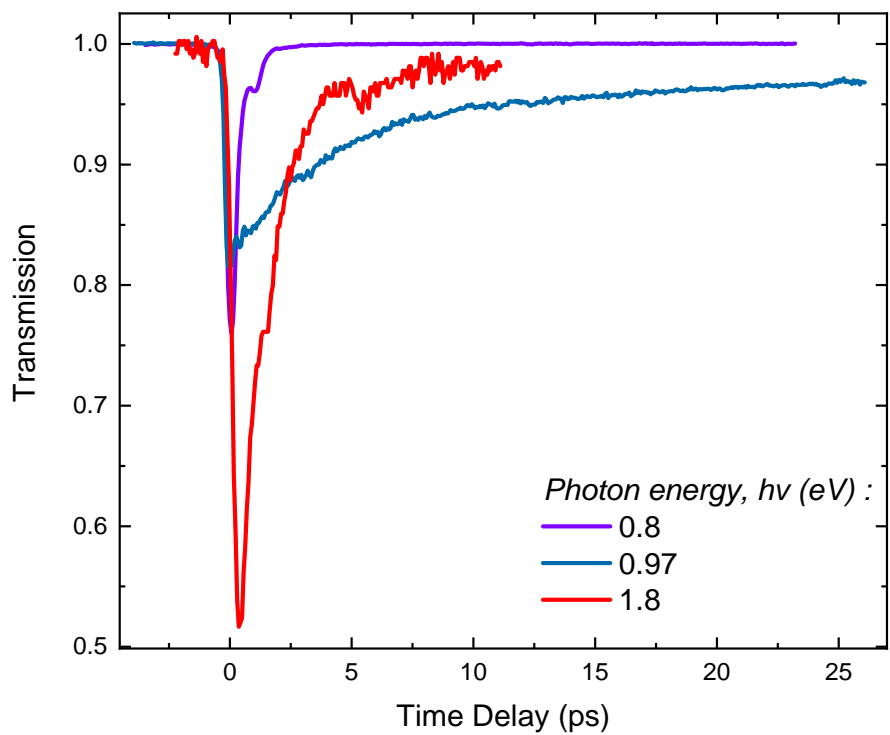

**Figure S3.** Pump-induced changes in the transient transmission as a function of the delay time $\Delta t$ between the pump and probe pulses for pump pulses with the photon energy tuned in resonance with $^3A_{2g} \rightarrow {}^3T_{1g}$ ($h\nu$=1.8 eV), $^3A_{2g} \rightarrow {}^3T_{2g}$ ($h\nu$=0.97 eV) transitions and in the transparency window ($h\nu$=0.8 eV). No coherent oscillations are visible in the transient transmission signal.

## 4. Magnetic field dependence

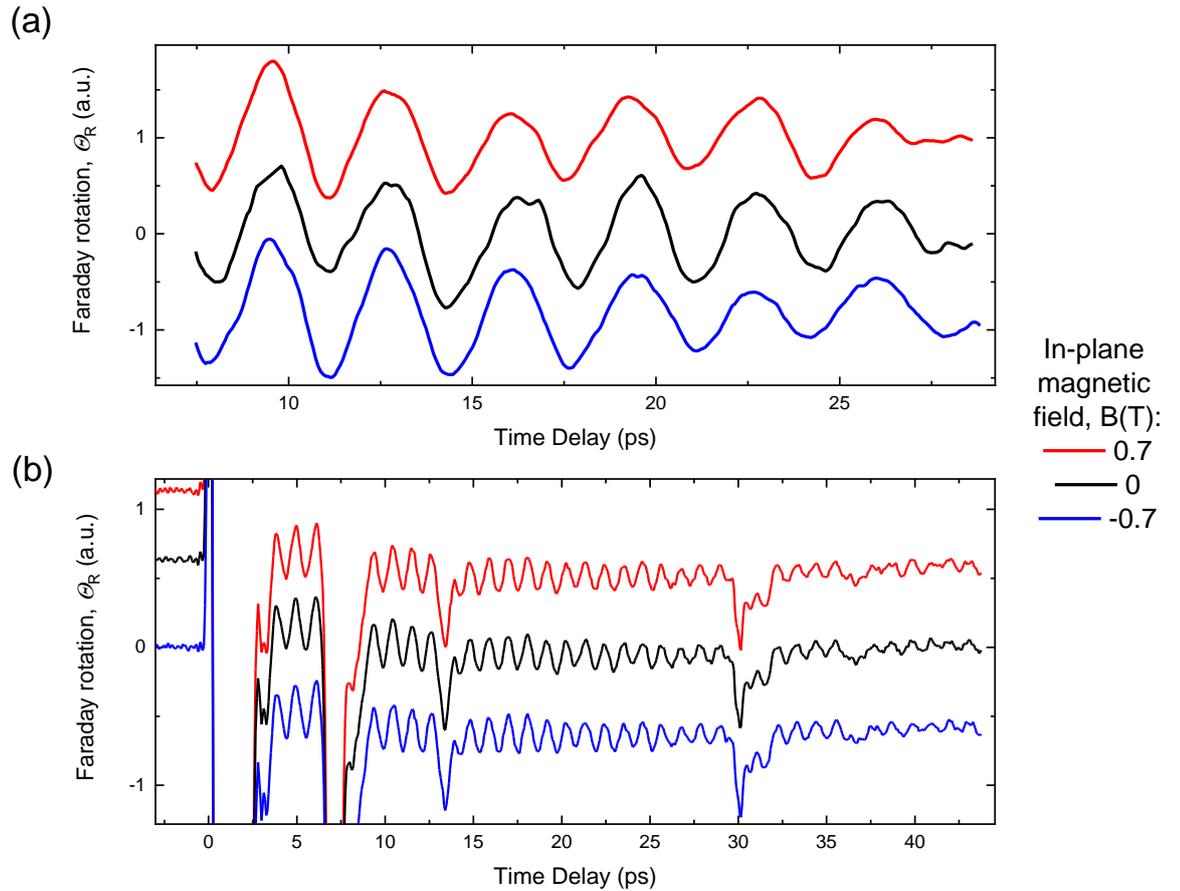

**Figure S4.** (a,b) Pump-induced traces of the polarization rotation plotted for different values of the external magnetic field applied in the sample *(ab)* plane, after excitation at a pump photon energy of 1 eV (a) and 0.2 eV (b). The data are shifted for clarity.

The application of magnetic fields, available in our laboratory, had no observable effect on the amplitude or frequency of the pump-induced oscillations.

## 5. Microscopic theory of the spin dynamics

The microscopic model Hamiltonian $\widehat{\mathcal{H}}$, describing the interaction of the antiferromagnetically coupled spins/chains is written as:

$$\widehat{\mathcal{H}} = \underbrace{\sum_{<ij>} J\widehat{S}_i\widehat{S}_j}_{\text{Exhange term}} + \underbrace{\sum_i (D_\perp \widehat{S}_{z,i}^2 + D_\parallel \widehat{S}_{y,i}^2)}_{\text{Biaxial magnetic anisotropy term}} - \underbrace{g\mu_B H \sum_i \widehat{S}_i}_{\text{Zeeman term}}$$

The sum $<ij>$ is carried out for all pairs of the nearest neighbours with the antiferromagnetic exchange interaction and each pair is considered only once.

$J > 0$ – phenomenological constant describing the strength of the antiferromagnetic exchange interaction between the neighbouring spins;

$D_\perp, D_\parallel > 0$ – phenomenological constants describing the strength of the out-of-plane and in-plane anisotropies, respectively. The form of the magnetic anisotropy term is selected, such that in the ground state the spins are oriented along the *x*-axis and can be obtained from a general form of the zero-field splitting (ZFS) Hamiltonian. It can be shown, see Appendix 5.1 below, that:

$$D_\perp = D + E \quad \text{and} \quad D_\parallel = 2E$$

where $D$ and $E$ are the axial and rhombic ZFS parameters, which account for distortions of the magnetic site along the main symmetry axis (*z*) and in the perpendicular *xy* plane, respectively.

$g$ – the value of the Landé factor for the $Ni^{2+}$ magnetic moment;

$\mu_B$ – Bohr magneton;

$H$ - external magnetic field.

The dynamics of the three components of the single spin-operator $\widehat{S}_i$ of the $i^{\text{th}}$ ion can be obtained via the Heisenberg equations of motion:

$$i\hbar \frac{d\widehat{S}_{1,2}}{dt} = [\widehat{S}_{1,2}, \widehat{\mathcal{H}}]$$

We perform a substitution $\widehat{S}_i \to S_i$, with $S_i$ a classical spin vector of fixed length to change from the exact quantum mechanical description of the problem to the semiclassical. The corresponding semiclassical (Landau-Lifshitz) equation of motion reads:

$$\hbar \frac{dS_{1,2}}{dt} = \left[S_{1,2} \times \frac{\partial W}{\partial S_{1,2}}\right]$$

with the magnetic energy:

$$W = zJS_1 S_2 + D_\perp (S_{1z}^2 + S_{2z}^2) + D_\parallel (S_{1y}^2 + S_{2y}^2) - g\mu_B (S_1 + S_2) H$$

$z$ – number of nearest neighbours with the antiparallel alignment of the spins;

We introduce transmutation combinations of classical spins $\mathbf{S}_{1,2}$ and $\mathbf{m}$ and $\mathbf{l}$ that are irreducible with respect to the group of sublattices, defined as:

$$\mathbf{S}_1 = S(\mathbf{m} + \mathbf{l})$$
$$\mathbf{S}_2 = S(\mathbf{m} - \mathbf{l}),$$

where $\mathbf{m}^2 + \mathbf{l}^2 = 1$, $\mathbf{ml} = 0$. $S(T)$ is the temperature-dependent averaged value of a single spin temperature behaviour of which in vicinity of $T_N$ typically described by a power law scaling such that:

$$S(T) \sim (T_N - T)^\beta$$

Where $\beta$ is the scaling power and $S \approx 1$ at low temperatures. The vectors $\mathbf{m}$ and $\mathbf{l}$ are normalized variants of the conventional Néel vector $\mathbf{L}$:

$$\mathbf{L} = S(\mathbf{S}_1 - \mathbf{S}_2)$$

and the net magnetization $\mathbf{M}$:

$$\mathbf{M} = S(\mathbf{S}_1 - \mathbf{S}_2).$$

The magnetic energy $W$ written in terms of $\mathbf{m}$ and $\mathbf{l}$ is:

$$W = 2zJS^2\mathbf{m}^2 + 2D_\perp S^2(l_z^2 + m_z^2) + 2D_\parallel S^2(l_y^2 + m_y^2) - 2g\mu_B H S m$$

The corresponding equations of motion for $\mathbf{m}$ and $\mathbf{l}$ are:

$$\begin{cases} 2\hbar S \dfrac{d\mathbf{m}}{dt} = \left[\mathbf{m} \times \dfrac{\partial W}{\partial \mathbf{m}}\right] + \left[\mathbf{l} \times \dfrac{\partial W}{\partial \mathbf{l}}\right] \\ 2\hbar \dfrac{d\mathbf{l}}{dt} = \left[\mathbf{m} \times \dfrac{\partial W}{\partial \mathbf{l}}\right] + \left[\mathbf{l} \times \dfrac{\partial W}{\partial \mathbf{m}}\right] \end{cases}$$

In the ground state, the Néel vector $\mathbf{l}$ is oriented along the $x$-axis, such that:

$$\mathbf{l} = l_0 \mathbf{e}_x \quad \mathbf{m} = 0$$

where $l_0 = \pm 1$ accounts for the time-reversal symmetry of the Néel state. As $\mathbf{m}^2 + \mathbf{l}^2 = 1$, one obtains $\mathbf{l}\partial \mathbf{l} = 0$ and hence the dynamics of the Néel vector is reduced to transverse deviations $\tilde{\mathbf{l}} = (\tilde{l}_y, \tilde{l}_z)$ of $\mathbf{l}$ from the equilibrium value:

$$\mathbf{l} = l_0 \mathbf{e}_x + \tilde{\mathbf{l}},$$

Here we consider small-amplitude oscillations of the Néel vector, which in the linear approximation leads to a pair of differential equations:

$$\begin{cases} 2\hbar S \dot{\tilde{\mathbf{m}}} = \left[l_0 \mathbf{e}_x \times \dfrac{\partial W}{\partial \mathbf{l}}\right] = 2l_0\{0, -2D_\perp S^2 l_z, 2D_\parallel S^2 l_y\} \\ 2\hbar S \dot{\tilde{\mathbf{l}}} = \left[l_0 \mathbf{e}_x \times \dfrac{\partial W}{\partial \tilde{\mathbf{m}}}\right] = 2l_0\{0, -2(zJ + D_\perp)S^2 m_z, 2(zJ + D_\parallel)S^2 m_y\} \end{cases} \quad (5.1)$$

These equations split into two pairs of linear differential equations for $(l_y, m_z)$ and $(l_z, m_y)$, the Cartesian components of the magnetic vectors $\boldsymbol{m}$ and $\boldsymbol{l}$:

$$\begin{cases} \hbar \dfrac{dm_z}{dt} = 2D_\parallel l_0 S l_y \\ \hbar \dfrac{dl_y}{dt} = -2(zJ + D_\perp) l_0 S m_z \end{cases} \qquad \begin{cases} \hbar \dfrac{dm_y}{dt} = -2l_0 D_\perp l_0 S l_z \\ \hbar \dfrac{dl_z}{dt} = 2(zJ + D_\parallel) l_0 S m_y \end{cases}$$

As a system of two first-order linear differential equations can be reduced to one 2$^{nd}$ order differential equation, we obtain two equations describing the in- and out-of-plane spin dynamics:

$$\ddot{l}_y + \left(\dfrac{l_0 S}{\hbar}\right)^2 4(zJ + D_\perp) D_\parallel l_y = 0 \qquad \ddot{l}_z + \left(\dfrac{l_0 S}{\hbar}\right)^2 4(zJ + D_\parallel) D_\perp l_z = 0$$

The corresponding frequencies of the in-plane and out-of-plane oscillatory motion are given:

$$\hbar \omega_\parallel = 2S\sqrt{D_\parallel(zJ + D_\perp)} \qquad \hbar \omega_\perp = 2S\sqrt{D_\perp(zJ + D_\parallel)}$$

As typically $zJ \gg D_{\perp,\parallel}$, the frequencies can be reasonably estimated by:

$$\hbar \omega_\parallel = 2S(T)\sqrt{D_\parallel(zJ)} \qquad \hbar \omega_\perp = 2S(T)\sqrt{(zJ)D_\perp}$$

One can see that the frequencies of the in-plane magnon and out-plane magnons are governed by the respective strength of the corresponding anisotropies. As the ZFS parameters define the magnetic anisotropies, the frequencies can be used to estimate $D$ and $E$ once the strength of the antiferromagnetic exchange $J$ is known.

Note, the frequencies are proportional to the average spin value $S(T)$ and thus have to be subjected to a power law scaling, same as the order parameter, i.e. the Neel vector **L**. The temperature dependence of the anisotropy constants $D_{\parallel,\perp}(T)$ can contribute to an additional dependence on $T$ can and the temperature variations of the anisotropy constant. As in our experiment $f_\parallel = f_1 \sim S(T)$ it clearly indicates that the mode is of the magnetic origin and that $D_\parallel(T)$ is independent of the temperature.

## 5.1 Appendix. Zero-field splitting Hamiltonian for a magnetic center with spin S

The general form of the one-spin zero-field splitting Hamiltonian reads:

$$\widehat{\mathcal{H}}_{\text{ZFS}} = \widehat{S} D \widehat{S}$$

with $D$ is diagonalizable matrix of zero-field crystal field parameters. In the diagonal form with the z-axis selected along the trigonal axis the Hamiltonian can be written as a triaxial Hamiltonian:

$$\widehat{\mathcal{H}}_{\text{ZFS}} = D\left(\hat{S}_z^2 - \frac{1}{3}S(S+1)\right) + E(\hat{S}_y^2 - \hat{S}_x^2)$$

where $D = \frac{3}{2} D_{zz}$ and $E = \frac{1}{2}(D_{yy} - D_{xx})$ are experimentally measurable values of the zero-field splitting. Here a positive value of $D$ is responsible for the easy-of-plane anisotropy due to the trigonal distortion and $E$ originates from the crystallographic inequivalence of the $x$ and $y$ axis due to a small rhombohedral distortion. It is thus also assumed that $|D| \gg |E|$.

Using $\hat{S}_x^2 + \hat{S}_y^2 + \hat{S}_z^2 = S(S+1)$ and neglecting spin-isotropic terms, the above Hamiltonian can be also written as an effective biaxial anisotropic Hamiltonian

$$\widehat{\mathcal{H}}_{\text{ZFS}} = D_\perp \hat{S}_z^2 + D_\parallel \hat{S}_y^2$$

where $D_\perp = D + E$ and $D_\parallel = 2E$

## 6. Relation between the spin-flop field $H_{sf}$ and the frequency of the in-plane magnon $\omega_\parallel$

We consider a magnetic field $\boldsymbol{H}$ applied along the *x*-direction:

$$W = 2zJS^2\boldsymbol{m}^2 + 2D_\perp S^2(l_z^2 + m_z^2) + 2D_\parallel S^2(l_y^2 + m_y^2) - 2g\mu_B H_x S m_x$$

In the magnetic *(xy)*-plane we introduce an angle $\theta$ which the vector $\boldsymbol{l}$ forms with the *x*-axis, such that:

$$\boldsymbol{m} = m(\boldsymbol{e}_x \cos\theta - \boldsymbol{e}_y \sin\theta);$$
$$\boldsymbol{l} = \sqrt{1-m^2}(\boldsymbol{e}_x \sin\theta + \boldsymbol{e}_y \cos\theta);$$

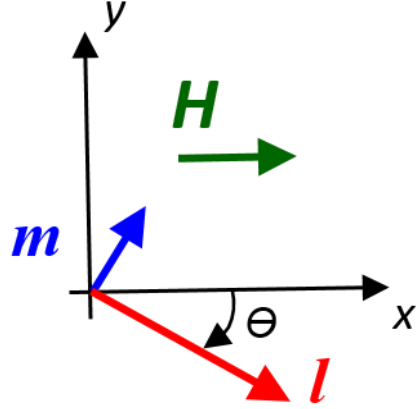

**Figure S5.** Spin-flop experimental geometry considered here.

The magnetic energy reads:

$$W = 2zJS^2 m^2 + 2D_\parallel S^2[(1-m^2)\sin^2\theta + m^2\cos^2\theta] - 2g\mu_B H_x S m \sin\theta$$

We first, minimize *W* with respect to *m*:

$$m(\theta) = \frac{g\mu_B H_x \sin\theta}{2(zJ + D_\parallel \cos 2\theta)S};$$

We now substitute $m(\theta)$ in the expression for *W*:

$$W(\theta) = -\frac{1}{2}\frac{(g\mu_B H_x)\sin^2\theta}{(zJ + D_\parallel \cos 2\theta)S} + D_\parallel S^2 \sin^2\theta\,;$$

and minimize with respect to $\theta$. Depending on the strength of the magnetic field $H_x$ two different solutions exist:

1. At $H_x < H_1 = \frac{2S\sqrt{D_\parallel(zJ+D_\perp)}}{g\mu_B}$ collinear AFM phase is stable and $\theta = 0 \quad m_x = 0$.

2. At $H_x > H_2 = \dfrac{S\sqrt{D_\parallel(zJ-D_\parallel)}}{g\mu_B}$ canted AFM phase is stable and $\theta = \dfrac{\pi}{2}$  $m_x = \dfrac{g\mu_B H_x}{2(zJ-D_\parallel)S}$.

Note, $H_1 > H_2$ and at $H_2 < H_x < H_1$ coexistence of the canted and collinear phases occurs.

Importantly, the higher field $H_1$ at which the collinear AFM phase loses its stability is directly related to the frequency of the in-plane spin precession:

$$H_1 = \dfrac{\hbar\omega_\parallel}{g\mu_B}$$

And thus, can serve to estimate its value:

$$\hbar\omega_\parallel = g\mu_B H_1$$

# 7. Light-matter interaction in NiPS$_3$ and the resulting spin dynamics

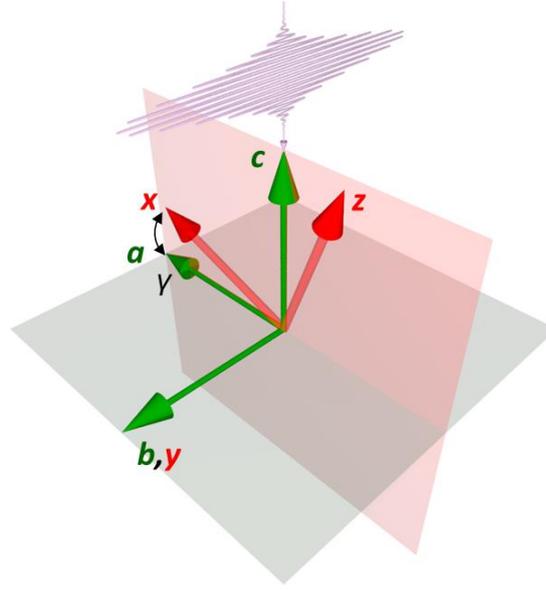

**Figure S6.** Geometry of the magnetic *(xyz)* and crystallographic *(abc)* reference planes with respect to the laser incidence.

To account for the observed interaction between the Ni$^{2+}$ spins and the ultrashort pulses we write the free energy term $W_{\text{int}}$ quadratic with respect to the electric field of light **E** and the antiferromagnetic vector *l*:

$$W_{\text{int}} = C_{ikpq} E_i E_k l_p l_q$$

which is allowed by the magnetic and crystallographic point group of NiPS$_3$. Here $C_{ikpq}$ are the components of the magnetoelectric susceptibility tensor *C*.

In the monoclinic system with C$_2$ symmetry axis along the *y*-axis, such as NiPS$_3$ the tensor *C* reads (in the Voigt notation):

$$C = \begin{pmatrix} C_{11} & C_{12} & C_{13} & C_{14} & 0 & 0 \\ C_{21} & C_{22} & C_{23} & C_{24} & 0 & 0 \\ C_{31} & C_{32} & C_{33} & C_{34} & 0 & 0 \\ C_{41} & C_{42} & C_{43} & C_{44} & 0 & 0 \\ 0 & 0 & 0 & 0 & C_{55} & C_{56} \\ 0 & 0 & 0 & 0 & C_{65} & C_{66} \end{pmatrix}, \begin{pmatrix} 1 \\ 2 \\ 3 \\ 4 \\ 5 \\ 6 \end{pmatrix} = \begin{pmatrix} xx \\ zz \\ yy \\ xz \\ yz \\ xy \end{pmatrix}$$

The electric field **E** of the laser pulse at the normal incidence to the surface of the sample (ab-plane) is $\mathbf{E} = (E_a \ E_b \ 0)$. Being projected on the *(xyz)* magnetic reference frame the vector **E** reads: $\mathbf{E} = (E_x \ E_y \ E_z) = (E_a \cos\gamma \ E_b \ E_a \sin\gamma)$ where $\gamma$ is an angle the *x*-axis forms with the *a*-axis in the *ac* plane.

Considering that in the linear approximation $l_x l_y$ and $l_x l_z$ are the only non-zero products of the Néel vector components which can contribute to the excitation of the spin dynamics we find:

$$W_{\text{int}} = (AE_a^2 + BE_b^2)l_x l_z + GE_a E_b l_x l_y,$$

where
$A = C_{14} \cos^2 \gamma + C_{24} \sin^2 \gamma + C_{44} \sin \gamma \cos \gamma;$
$B = C_{34};$
$G = C_{56} \sin \gamma + C_{66} \cos \gamma.$

In our experiments, the laser light is linearly polarized with the polarization plane oriented at angle polarization $\phi$ oriented along the $a$-axis and therefore $\mathbf{E} = (E_0 \cos \phi \quad E_0 \sin \phi \quad 0)$, where $E_0(t) \approx E_0 \delta(t)$ is a time-dependent component of the electric field strength of the laser pulse. Thus, the interaction energy reduces to:

$$W_{\text{int}} = \frac{1}{2} E_0^2(t) \big[ ((A + B) + (A - B) \cos 2\phi) l_x l_z + G \sin 2\phi \, l_x l_y \big]$$

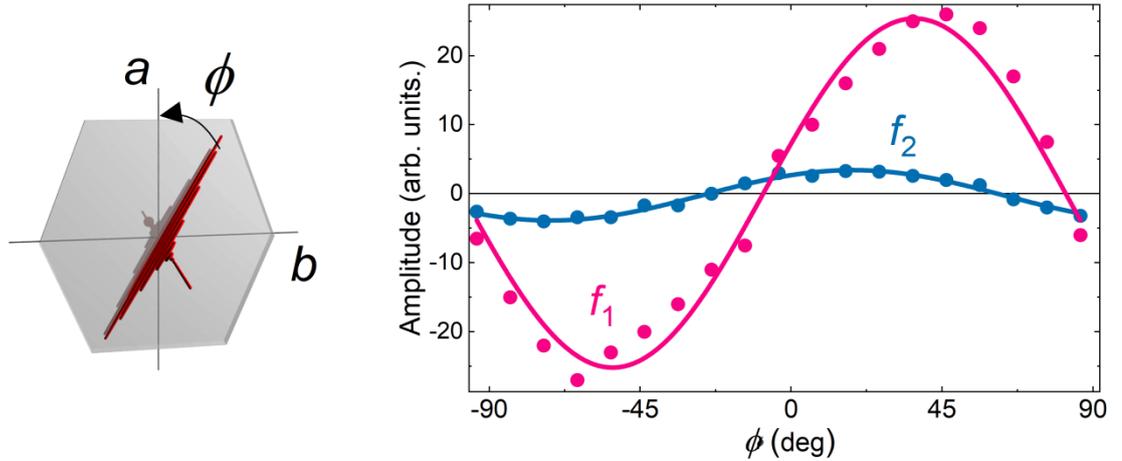

**Figure S7.** The amplitude of the oscillations at $f_1$ and $f_2$ as a function of the azimuthal angle $\phi$ between the pump polarization plane and the $a$-crystal axis. Schematics showing the geometry of the experiment.

The $W_{\text{int}}$ is anisotropic with respect to the components of the $\mathbf{l}$ vector and thus within presence of the light in the media acts on the spins as an effective magnetic anisotropy. This action generates a torque $\frac{\partial W_{\text{int}}}{\partial l}$ which leads to a time-dependent force term $F_{y,z}(t)$ on the right-hand side of the equations of the motion (Eq. 5.1):

$$\begin{cases} 2S\hbar \dot{m}_z = l_0(2D_\parallel S^2 l_y + F_y(t)l_0) \\ \hbar \dot{l}_y = -2(zJ + D_\perp)l_0 S m_z \end{cases} \qquad \begin{cases} 2S\hbar \dot{m}_y = -l_0(2D_\perp S^2 l_z + F_z(t)l_0) \\ \hbar \dot{l}_z = 2(zJ + D_\parallel)l_0 S m_y \end{cases}$$

$$F_y(t) = G \sin 2\phi \cdot E_0^2(t) \qquad F_z(t) = \frac{1}{2}((A+B) + (A-B)\cos 2\phi)E_0^2(t)$$

The corresponding equations of the spin motion read:

$$\ddot{l}_y + \omega_\parallel^2 l_y = -\frac{(zJ + D_\perp)l_0}{2\hbar^2} F_y(t)$$

$$\ddot{l}_z + \omega_\perp^2 l_z = -\frac{(zJ + D_\parallel)l_0}{2\hbar^2} F_z(t)$$

Considering that duration of the pump pulse $\Delta t$ (~100 fs) than the period of spin oscillations, $\Delta t \ll \frac{2\pi}{\omega_{\parallel,\perp}}$, the real pulse shape can be replaced by the Dirac delta function $\delta(t)$: $E_0^2(t) \to I_0 \delta(t)$ where $I_0 = \int E_0^2(t) dt$ is the integrated pulse intensity. This substitution clearly shows that action of the pump pulse is reduced to an instantaneous force/torque impulsively launching the spin dynamics.

One can see that to excite the in-plane dynamics $l_y$ of the Néel vector, the polarization of light has to be oriented off the *a* and *b* axis of the crystal and reaches maxima of opposite sign at $\phi = \pm 45$ degrees in perfect agreement with our experiment and the ultrafast Inverse Cotton-Mouton effect.

Interestingly, the formulas also clearly show that excitation of the out-of-plane dynamics $l_z$ of the Néel vector is also possible at the normal incidence. Note, the excitation is allowed only because of the monoclinic distortion as the excitation relies on the non-zero components A, B, each depending on the monoclinic terms $C_{i4}$ of the magnetoelectric susceptibility tensor *C*.

Remarkably, the dependence of the magnon amplitude on the orientation of the pump pulse is expected to be the same as for the in-plane mode but 45 degrees shifted (compare $\sin 2\phi$ and $\cos 2\phi$). In the experiment, a 180 degrees dependence of the coherent oscillation at $f_2$ was also observed. However, only a 22 degrees shift was observed as compared to the $f_1$ mode, see Fig. S7. The discrepancy can be explained by the birefringence of the crystal which influences the linear polarization of the incident pump light.